\documentstyle[prl,aps]{revtex}
\draft
\begin{document}
\title{Potential Scattering and the Kondo Effect}
\author{Valery I. Rupasov}
\address{Department of Physics, University of Toronto, Toronto,
Ontario, Canada M5S 1A7\\
and Landau Institute for Theoretical Physics, Moscow, Russia}
\date{\today}
\maketitle
\begin{abstract}
We study a generalized Kondo model in which a spin-$\frac{1}{2}$
impurity is coupled to a conduction band by both s-d exchange
and potential interactions. A strong potential scattering
is shown to screen an exchange scattering, and the Kondo
temperature of the system is decreased.
\end{abstract}


It is well known that the behavior of a magnetic impurity
in a metal can be described by an effective 1D Hamiltonian,
\begin{equation}
H=\sum_{\sigma}\int\frac{dk}{2\pi}k
c^\dagger_\sigma(k)c_\sigma(k)
+\frac{1}{2}I\sum_{\sigma,\sigma'}
\int\frac{dk}{2\pi}\frac{dk'}{2\pi}c^\dagger_\sigma(k)
\left(\vec{\sigma}_{\sigma\sigma'}\cdot\vec{S}\right)
c_{\sigma'}(k')
+V\sum_{\sigma}\int\frac{dk}{2\pi}
\frac{d\epsilon'}{2\pi}
c^\dagger_\sigma(k)c_{\sigma}(k').
\end{equation}
Here, $\vec{\sigma}$ are the Pauli matrices, $\vec{S}$ is
the impurity spin operator, and $I$ is the s-d exchange
coupling constant. The operators $c^\dagger_\sigma(k)$
refer to conduction electrons with spin
$\sigma=\uparrow,\downarrow$ in a $s$-wave state of
momentum modulus $k$. The last term in Eq. (1) corresponds
to a potential (spin independent) scattering of an electron
on the impurity site, $V$ being the coupling constant.
The electron energies and momenta in Eq. (1) are taken
relative to the Fermi values, which are set to be equal
to zero, while the Fermi velocity $v_F=1$. For simplicity,
we confine ourselves to the case of $S=1/2$ which is of
the most physical interest.

In the absence of a potential coupling, $V=0$, the model
(1) is solved exactly \cite{W,A} by the Bethe ansatz (BA).
In the BA approach to the theory of dilute magnetic alloys
\cite{TW,AFL}, the spectrum of a host is alternatively
described in terms of charge and spin excitations rather
than in terms of free particles with spin ``up'' and `down''.
The BA equations for charge, $k_j$, $j=1,\ldots,N$, and spin
$\lambda_\alpha$, $\alpha=1,\ldots,M$, rapidities describing
$N$ particles on an interval of size $L$ have the form \cite{N}
\begin{mathletters}
\begin{eqnarray}
\exp{(ik_jL)}\,\Phi_{\mbox{ch}}&=&\left(\frac{\lambda_\alpha+\frac{i}{2}}
{\lambda_\alpha-\frac{i}{2}}\right)^M\\
\left(\frac{\lambda_\alpha+\frac{i}{2}}{\lambda_\alpha-\frac{i}{2}}\right)^N
\Phi_{\mbox{sp}}(\lambda_\alpha)&=&-\prod_{\beta=1}^{M}
\frac{\lambda_\alpha-\lambda_\beta+i}{\lambda_\alpha-\lambda_\beta-i}
\end{eqnarray}
where $M$ is the number of particles with spin ``down''.
The eigenenergy $E$ and the $z$ component of the total
spin of the system $S^z$ are found to be
\begin{equation}
E=\sum_{j=1}^{N}k_j,\;\;\;S^z=\frac{1}{2}+\frac{N}{2}-M.
\end{equation}
In Eqs. (2) the phase factors
\end{mathletters}
\begin{mathletters}
\begin{eqnarray}
\Phi_{\mbox{ch}}&=&\frac{1-\frac{i}{2}U_0}{1+\frac{i}{2}U_0}\simeq
\exp{(-iU_0)}\\
\Phi_{\mbox{sp}}(\lambda)&=&
\frac{\lambda+\frac{1}{\mbox{g}_0}+\frac{i}{2}}
{\lambda+\frac{1}{\mbox{g}_0}-\frac{i}{2}}
\end{eqnarray}
describe the scattering of charge and spin excitation
of the host on the impurity. Here
\begin{equation}
U_0=\frac{1}{4}I,\;\;\; \mbox{g}_0=\frac{1}{2}I.
\end{equation}

In the absence of the exchange coupling, $I=0$ or $S=0$,
multiparticle effects are also absent. The model (1) is
diagonalized in terms of independent particles with
spin ``up'' and ``down'' scattering on the impurity
potential. Pure potential impurities are clear to change
physical properties of a host only if they significantly
modify the density of band states near the Fermi level.
Therefore, dramatic changes in the thermodynamics of a
host under a doping with magnetic impurities is associated
with an s-d exchange coupling only, while the potential
scattering term is assumed can be omitted.

Here, we note that not changing the structure
of the BA equations (2) a potential scattering renormalizes
the parameters $U_0$ and $\mbox{g}_0$. In the presence of
potential scattering these parameters are replaced,
respectively, by
\end{mathletters}
\begin{mathletters}
\begin{eqnarray}
U&=&U_0+V\\
\mbox{g}&=&\frac{\mbox{g}_0}
{1+\frac{1}{4}(V+\frac{1}{4}I)(V-\frac{3}{4}I)}.
\end{eqnarray}

At $V=0$, Eqs. (4) reduce to expressions given in
Eq. (3c), provided that, as it is assumed in deriving
Eqs. (2), $I\ll 1$. In the domain $I\ll |V|\ll 1$, the
impurity term in Eq. (2a) for charge excitation is
determined by potential scattering, $U\simeq V$, while
scattering of spin excitations at the impurity is still
determined by exchange coupling, $\mbox{g}\simeq\frac{1}{2}I$.
Finally, at $|V|\gg 1$, the effective coupling constant
is given by $\mbox{g}=2I/V^2=4\mbox{g}_0/V^2$, and
hence the exchange scattering is screened by the potential
one.

Thus, a strong potential scattering does not affect
the qualitative behavior of the system, but it renormalizes
physical parameters of the Kondo effect. In particular,
the Kondo temperature is now found to be
\end{mathletters}
\begin{equation}
T_K\sim\epsilon_F\exp{\left(-\frac{\pi}{\mbox{g}}\right)}
=\epsilon_F\exp{\left(-\frac{\pi V^2}{\mbox{2I}}\right)},
\end{equation}
where $\epsilon_F$ is the Fermi energy, while in the absence
of potential scattering $T_K^0\sim\epsilon_F\exp{(-2\pi/I)}$.

To derive Eqs. (4), it is enough to study how a potential
scattering is incorporated in the particle-impurity scattering
matrix. Let us look for one-particle eigenstates of the system
in the form
$$
|\Psi_1\rangle=\sum_{\sigma}\sum_{s=0,1}\int\frac{dk}{2\pi}
\psi_{\sigma;s}(k)c^\dagger_\sigma(k)
\left(S^+\right)^s|0\rangle,
$$
where the vacuum state $|0\rangle$ contains no electrons
and the impurity spin is assumed to be ``down''. The
Schr\"odinger equation is then easily found to be
\begin{mathletters}
\begin{eqnarray}
&&(k-\omega)\psi(k)+\frac{1}{2}I\left(\vec{\sigma}\cdot\vec{S}\right)
A+VA=0\\
&&A=\int\frac{dk}{2\pi}\psi(k),
\end{eqnarray}
where $\omega$ is the eigenenergy, and spin indexes are omitted.

Inserting the general solution of Eq. (6a),
\end{mathletters}
$$
\psi(k)=2\pi\delta(k-\omega)\chi-\frac{V}{k-\omega-i0}A
-\frac{\frac{1}{2}I}{k-\omega-i0}
\left(\vec{\sigma}\cdot\vec{S}\right)A,
$$
with an arbitrary spinor $\chi$, into Eq. (6b), one
obtains
$$
\left\{1+\frac{i}{2}\left[V+
\frac{1}{2}I\left(\vec{\sigma}\cdot\vec{S}\right)\right]\right\}A=
\chi.
$$
For the Fourier image of the function $\psi(k)$,
$$
\psi(x)=\int_{-\infty}^{\infty}\frac{dk}{2\pi}
\psi(k)\exp{(ikx)},
$$
a solution of Eqs. (6) is then found to be
$$
\psi(x)=e^{ikx}\left\{
\begin{array}{rl}
\chi,&x<0\\
{\bf R}\chi,&x>0
\end{array}
\right.
$$
where $k=\omega$. Here, the electron-impurity scattering matrix
is given by
\begin{mathletters}
\begin{equation}
{\bf R}=u+2v\left(\vec{\sigma}\cdot\vec{S}\right),
\end{equation}
where the parameters $u$ and $v$ are found from the equations
\begin{eqnarray}
u+v&=&\frac{1-\frac{i}{2}(V+\frac{1}{4}I)}
{1+\frac{i}{2}(V+\frac{1}{4}I)}\\
u - 3v&=&\frac{1-\frac{i}{2}(V-\frac{3}{4}I)}
{1+\frac{i}{2}(V-\frac{3}{4}I)}.
\end{eqnarray}

Diagonalization of the system with the particle-im\-pu\-ri\-ty
scattering matrix ${\bf R}$ results in the BA equations
(2) with scattering parameters given in Eqs. (4).

A screening of the exchange scattering and a lowering the
Kondo temperature by quite a strong potential scattering,
$|V|\gg I$, find a natural physical explanation. In the
absence of the potential scattering, multiparticle effects
in a spin subsystem of a metal doped with a Kondo impurities
are generated due to an essential difference of electron-impurity
scattering amplitudes in the triplet, $u+v$, and singlet,
$u- 3v$, channels of scattering. As $V$ grows, this difference
in scattering amplitudes is easily seen from Eqs. (7) to be
decreased and become negligible small at $|V|\gg I$. Therefore,
the impurity term $\Phi_{\mbox{sp}}$ disappears from Eq. (2b)
describing the behavior of the spin subsystem, and the impurity
is decoupled from the band states.

\end{mathletters}


\begin{references}
\bibitem{N}
We follow notation used in the review paper of Tsvelick
and Wiegmann \cite{TW}.
\end{references}
\end{document}